\documentclass{aa}
\usepackage{graphicx}
\usepackage{natbib}
\bibpunct{(}{)}{;}{a}{}{,} %
\normalfont
\newcommand{\kms}{\rm ~km~s^{-1}}
\newcommand{\ccm}{\rm ~cm^{-3}}

\newcommand{\ang}{\mbox{ \AA}}
\newcommand{\etal}{\rm et al.~}
\def\EE#1{\times 10^{#1}}

\def\cm2{\rm ~cm^{-2}}
\def\cm3{\rm ~cm^{-3}}
\def\kms{\rm ~km~s^{-1}}

\def\ergs{\rm ~erg~s^{-1}}
\def\ergsm{\rm ~erg~s^{-1} cm^{-2}}
\def\wl{~\lambda}
\def\wll{~\lambda\lambda}

\def\Ti44{M(^{44}{\rm Ti})}

\def\lsim{\!\!\!\phantom{\le}\smash{\buildrel{}\over
  {\lower2.5dd\hbox{$\buildrel{\lower2dd\hbox{$\displaystyle<$}}\over
                               \sim$}}}\,\,}
\def\gsim{\!\!\!\phantom{\ge}\smash{\buildrel{}\over
  {\lower2.5dd\hbox{$\buildrel{\lower2dd\hbox{$\displaystyle>$}}\over
                               \sim$}}}\,\,}

\begin{document}
   \title{Coronal emission from the shocked
   circumstellar ring of SN 1987A \thanks{Based on 
   observations performed at the European Southern Observatory, Paranal, 
   Chile.}}


 \author{Per Gr\"oningsson\inst{1}, 
Claes Fransson\inst{1}, Peter
Lundqvist\inst{1}, Tanja Nymark\inst{1}, Natalia
Lundqvist\inst{1}, Roger Chevalier\inst{2},
Bruno Leibundgut\inst{3}, Jason Spyromilio\inst{3}
}

   \offprints{P. Gr\"oningsson}

   \institute{
	Department of Astronomy, Stockholm University,
              AlbaNova University Center, SE-106 91 Stockholm,
	Sweden, \and Department of Astronomy, University of
Virginia, P.O. Box 3818, Charlottesville, VA 22903,\and
European Southern Observatory,
Karl-Schwarzschild-Strasse 2, D-85748 Garching, Germany.\\
              \email{per@astro.su.se
}
	}

   \date{Received ; accepted }

\titlerunning{Coronal emission from SN~1987A}
\authorrunning{Gr\"oningsson et al.}

\abstract{High resolution spectra with UVES/VLT of SN 1987A from
December 2000 until November 2005 show a number of high ionization
lines from gas with velocities of $\sim \pm 350 \kms$, emerging from
the shocked gas formed by the ejecta -- ring collision. These include
coronal lines from [Fe~X], [Fe~XI] and [Fe~XIV] which have
increased by a factor of  $\sim 20$ during the observed
period. The evolution of the lines is similar to that of the soft
X-rays, indicating that they arise in the same component. The line
ratios are consistent with those expected from radiative shocks with
velocity $310-390 \kms$, corresponding to a shock temperature of
$(1.6-2.5)\EE6$ K. A fraction of the coronal emission may, however,
originate in higher velocity adiabatic shocks.

   }

   \maketitle
	
%

\section{Introduction}
The collision of the ejecta of SN 1987A with the circumstellar ring
has been observed in many wavelength ranges (e.g.,
\cite{mccray05} for a review). In the UV/optical range this has been
monitored with HST showing an increasing number of hot spots around the ring
\citep{michael00,michael02,pun02}. Most likely, these hot spots are
caused by the impact of the blast wave on protrusions from the inner 
circumstellar ring. As time progresses an increasing number of these are 
expected to appear, and finally the whole ring will be immersed in the 
collision. 

The spectroscopic HST observations have identified three different velocity
components \citep{pun02}. One narrow, nearly Gaussian, with FWHM $\sim
10 \kms$, coming from the unshocked ring gas. Secondly, there is an
intermediate velocity component arising from shocked cooling gas, with
a clearly non-Gaussian form, extending to $\sim \pm 300 \kms$. Finally
there is a very broad component extending to $\sim 15\,000 \kms$,
probably coming from the reverse shock, resulting from the interaction
with the surrounding medium \citep{michael98,smith05}. We will refer to
these as the narrow, intermediate and high velocity components,
respectively. 

The collision is also seen in the radio as a rising non-thermal
synchrotron flux \citep{gaensler97,man02,man05}. The light curves in
this range show a roughly linear increase with time.
Further, \cite{bouchet06} find from ground based and Spitzer mid-IR
observations evidence for hot dust as well as highly ionized line
emission.   

In X-rays ROSAT
and Chandra have observed SN 1987A at several epochs, and also in this
energy range a steadily rising flux is seen
\citep{Park04,Park05}. While the hard X-rays correlate well with the
radio flux, the rise time in the soft X-rays is considerably
shorter. Of special interest are the high resolution spectral
observations \citep{Zhe05,zhe06}, which show a number of lines from H and He
like ions of N, O, Ne, Si, Mg and S, as well as Fe XVII. \cite{Zhe05}
argue that these lines arise in the radiative shocks propagating into the
protrusions, as well as the reflected shock propagating back from
these. In this paper we discuss optical observations complementary
to these X-ray observations of highly ionized gas.

In \citet{gro06} we report on spectral 
shapes and fluxes of optical lines created by the ejecta/ring 
interaction. As we use high-resolution echelle spectroscopy,
we can easily disentangle the different velocity components from each other. 
Here we concentrate on the evolution of intermediate velocity high-ionization  
lines, such as [Fe~X]~$\lambda$6374.5, [Fe~XI]~$\lambda$7891.8 and 
[Fe~XIV]~$\lambda$5302.9, as they are efficient probes of the shocked
gas. 

In our analysis we use a recession velocity of $281 \kms$ (the bulk
motion of the northern side of the inner circumstellar ring
\citep{gro06}) and a reddening of $E_{B-V}=0.16$ \citep{FW90} with
$E_{B-V}=0.06$ from the Milky way \citep[e.g.,][]{SS03} and $E_{B-V}=0.10$ from the LMC. The
reddening law was taken from \citet{Card89} using
$R_{V}=3.1$. The differences between the LMC extinction law and the
Galactic law are negligible in the optical at low color excess
\citep{F99} and have therefore been ignored.  In Sect. 2 and 3
we describe the observations and observational results,
respectively. In Sect. 4 we discuss modeling of the lines, followed by
a general discussion in Sect. 5 and conclusions in Sect. 6.


\section{Observations}
SN 1987A was observed in service mode with the Ultraviolet and Visual
Echelle Spectrograph (UVES) at ESO/VLT at Paranal, Chile.  UVES is a
cross-dispersed echelle spectrograph covering (with two different
dichroic settings) the spectral range $\sim 3100-10000\ang$
\citep{dekker00}. The light beam is split up in two separate arms. The
arm covering the blue part of the spectrum ($\sim 3100-4900\ang$) has
a single CCD detector with a spatial resolution of $0\farcs246/{\rm
pix}$. The other arm covering the red part ($\sim 4800-10000 \ang$) is
equipped with two CCDs having the somewhat higher spatial resolution
of $0\farcs182/{\rm pix}$. The slit width used resulted in a resolving
power of $\sim 50\,000$ corresponding to a spectral resolution of
$\sim 6 \kms$.

\begin{table*}
\centering
\caption{Log of UVES observations.}
\begin{tabular}{l l c c c c c  c c}
\hline
\hline
Epoch&Date&Days after&Grating&$\lambda$ range&Slit
width&Resolution&Exposure&Seeing\\ 
&&explosion&&\AA&arcsec&$\lambda/\Delta\lambda$&s&arcsec\\
\hline
1&1999 Oct 16&4618&DIC1 346+580&3100--3900&1.0&40,000&1,200&1.0\\
&&&&4800--6800&&&&\\
&&&&&&&&\\
2&2000 Dec 9--14&5039--5043&DIC1 346+580&3100--3900&0.8&50,000&10,200&0.4--0.8\\
&&&&4800--6800&&&&\\
&&5038--5039&DIC2 390+860&3300--4400&&&9,360&0.4\\
&&&&6700--10000&&&&\\
&&&&&&&&\\
3&2002 Oct 4--7&5704--5705&DIC1 346+580&3100--3900&0.8&50,000&10,200&0.7--1.0\\
&&&&4800--6800&&&&\\
&&5702--5703&DIC2 437+860&3800--4900&&&9,360&0.4--1.1\\
&&&&6700--10000&&&&\\
&&&&&&&&\\
4&2005 Mar 21&6601--6623&DIC1 346+580&3100--3900&0.8&50,000&9,200&0.6--0.9\\
&Apr 8-12&&&4800--6800&&&&\\
&&6621--6623&DIC2 437+860&3800--4900&&&4,600&0.5\\
&&&&6700--10000&&&&\\
&&&&&&&&\\
5&2005 Oct 20&6826&DIC1 346+580&3100--3900&0.8&50,000&2,300&0.9\\
&Nov 1--12&&&4800--6800&&&\\
&&6814--6837&DIC2 437+860&3800--4900&&&9,200&0.5--1.0\\
&&&&6700--10000&&&&\\

\hline

\hline
\label{tab:log}
\end{tabular}

\end{table*}

Spectra of the ring and ejecta of SN 1987A were obtained on October 16
1999, December 10--14 2000, October 4--7 2002, March 21 -- April 12
2005 and October 20 -- November 12 2005, in the following referred to as
epochs 1--5, respectively.  In all cases, except for epoch 1, the
position angle was PA$=30^{\circ}$.  For epoch 1 it was
PA$=20^{\circ}$.  The log of the observations is given in Table
\ref{tab:log}. The observations and data reduction are discussed in
detail in \cite{gro06}. As can be seen from Table \ref{tab:log}, the
exposure of the first epoch was very short and the S/N low. This made
it impossible to obtain meaningful fluxes for the coronal lines for
this epoch, and therefore we do not discuss it further in this paper.

Because the spatial resolution of these ground-based observations is
limited compared to the dimensions of the ring, we cannot
distinguish between different hot spots located on the same side of
the ring. To identify how many hot spots were
covered by the slit we retrieved and studied HST images taken with the 
WFPC2 and ACS instruments at roughly the same epochs. These show that
only Spot 1 is responsible for the emission from the shocked ring at
epoch 1. At epoch 2 Spot 1 still dominates the shocked gas emission on 
the north side of the ring, but also two spots on the opposite side of 
the ring are now prominent. At epoch 3, three to four different hot spots 
contribute to the emission on the northern side and by the time of the last 
epochs the entire ring is lit up by the ejecta-ring collision. Because
of the difference in velocity and to isolate a limited part of the
ring we will in the following only discuss the spectrum from the north side
of the ring, i.e., where Spot 1 first appeared. A comparison of the
the kinematics and flux of the two sides will be given in \cite{gro06}.

To estimate the accuracy of the absolute flux level of the spectra we
compared the individual sensitivity functions created from the reduced
standard star spectra and their corresponding physical fluxes. The
sensitivity curves differed typically by $\lsim 10\%$ from each
other. These exposures of the standard stars used a wide slit
($10\arcsec$) and hence no slit losses are likely to occur. Our
science data exposures, on the other hand, used a relatively narrow
slit and since SN 1987A is an extended source the slit losses could be
considerable and depend on atmospheric conditions such as the
seeing. To estimate how much the atmospheric conditions could
influence our results we have compared line fluxes of the individual
science spectra for the epochs which have more than one exposure. This
comparison reveals that the line fluxes differ by $\lsim 10\%$ from
one exposure to another. Finally, to estimate the total systematic
error, we have compared our fluxes with HST spectra and photometry
\citep[see][]{gro06}, and find that the accuracy of the VLT fluxes
should in general be better than $\sim 30\%$.

\section{Results}
\label{sec_res}

In \cite{gro06} we give a full list of the different lines present,
both from the unshocked and the shocked ring.  The former are
dominated by thermal broadening with a FWHM of $\sim 10 \kms$. The
lines from the shocked gas are on the other hand dominated by the
macroscopic velocity and are highly asymmetrical, extending to $\sim
250-400 \kms$ (HWZI). The two components can therefore easily be well
separated, and a subtraction of the broader from the narrower
component by interpolation of the intermediate components by
regression splines is fairly straightforward. More problematic is the
blending of different lines from the shocked component. In particular,
the large number of Fe II lines from this component causes a
considerable number of blendings.

The full spectrum between 3100-10000 \AA \ contains $\sim~170$
intermediate velocity lines. About one third of these are Fe II lines,
but there are also strong lines from more highly ionized ions, like
[O~III] $\wll 4958.9, 5006.8$, [Ne~III] $\wll 3868.9, 3967.5$, [Ne~V]
$\wll 3345.3, 3425.9$, and [Ar~III] $\wl 7135.8$. The most interesting
result is, however, that we find a number of highly ionized coronal
lines, which can only be seen thanks to the high S/N and high spectral
resolution of our UVES observations.

\begin{figure}[h]
\begin{center}
\resizebox{\hsize}{!}{\includegraphics{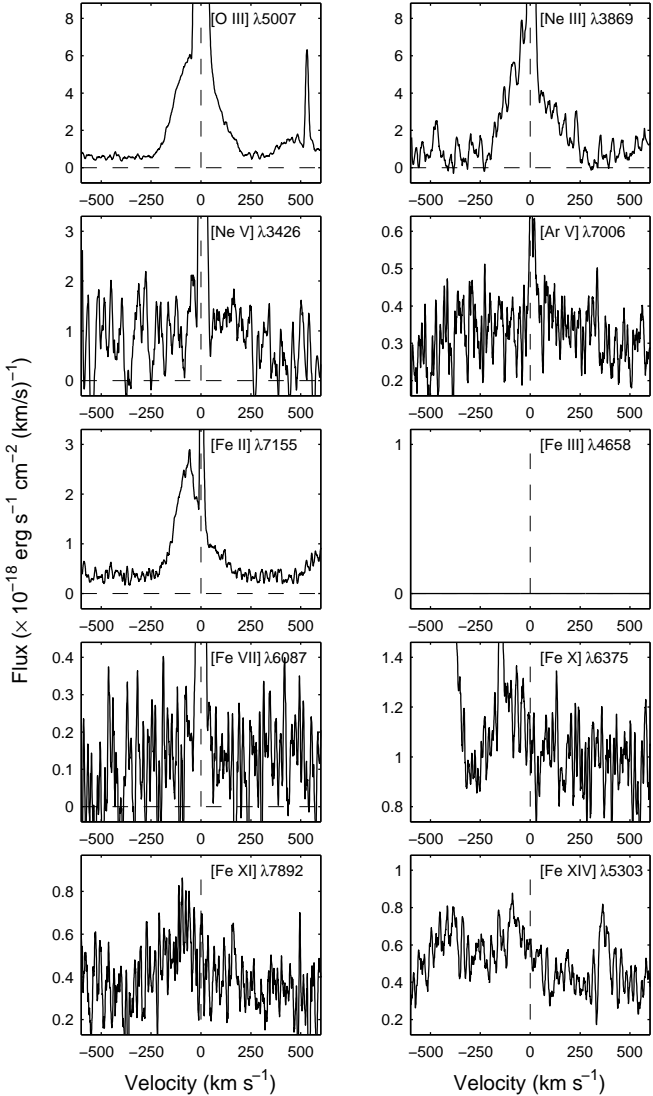}}
\end{center}
\caption{ Original line profiles from December 2000 (our epoch 2) of
the most important high ionization lines from the north side of the
shocked ring. The [Fe~III]$\wl 4658.1$ line was not covered by the
observations at this epoch. Note the different flux scales. All
velocities are with respect to the systemic velocity of the northern
part of the ring, $281 \kms$}.
\label{fig:high_ion_raw0012}
\end{figure}

\begin{figure}[h]
\begin{center}
\resizebox{\hsize}{!}{\includegraphics{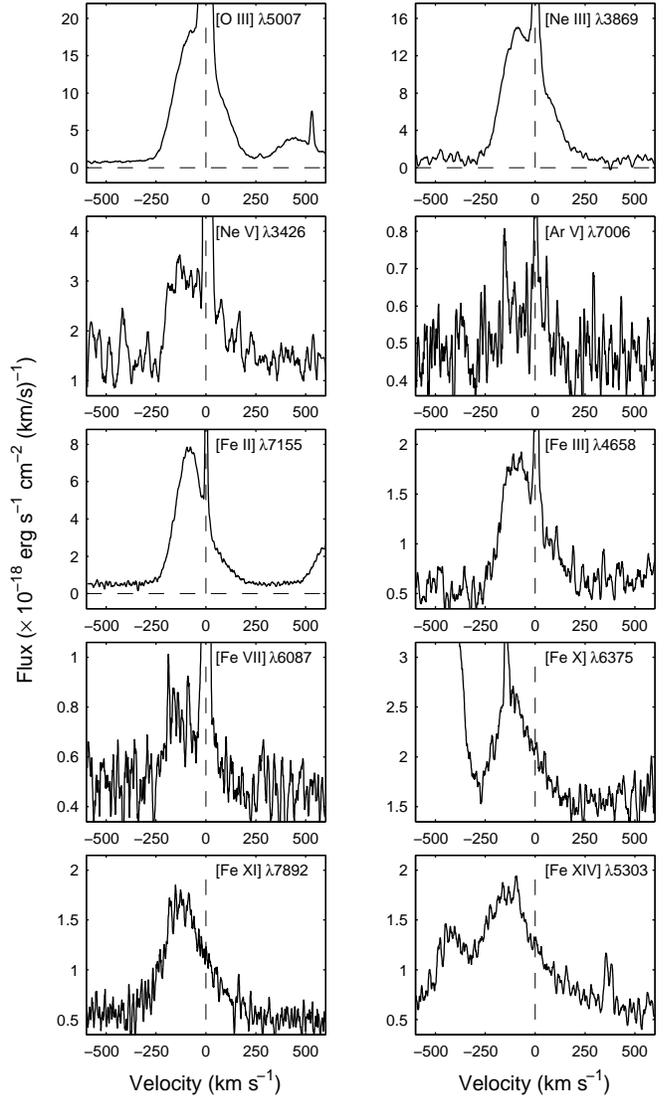}}
\end{center}
\caption{ Same as Fig. \ref{fig:high_ion_raw0012}, but for October 2002.}
\label{fig:high_ion_raw0210}
\end{figure}

\begin{figure}[h]
\begin{center}
\resizebox{\hsize}{!}{\includegraphics{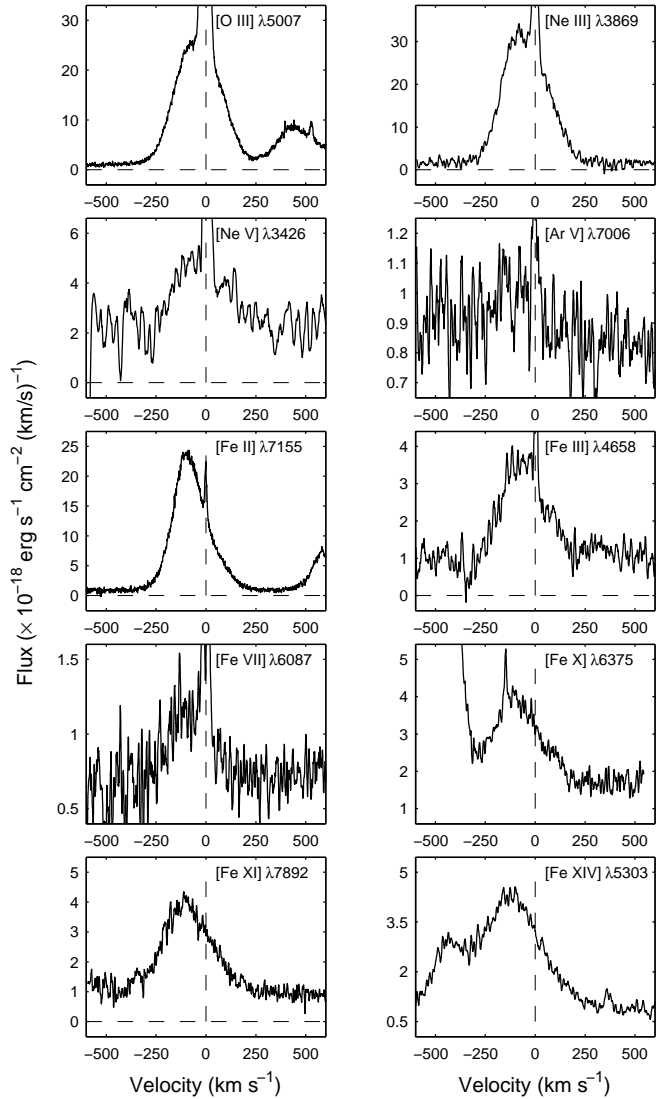}}
\end{center}
\caption{ Same as Fig. \ref{fig:high_ion_raw0012}, but for March -- April 2005.}
\label{fig:high_ion_raw0503}
\end{figure}

\begin{figure}[h]
\begin{center}
\resizebox{\hsize}{!}{\includegraphics{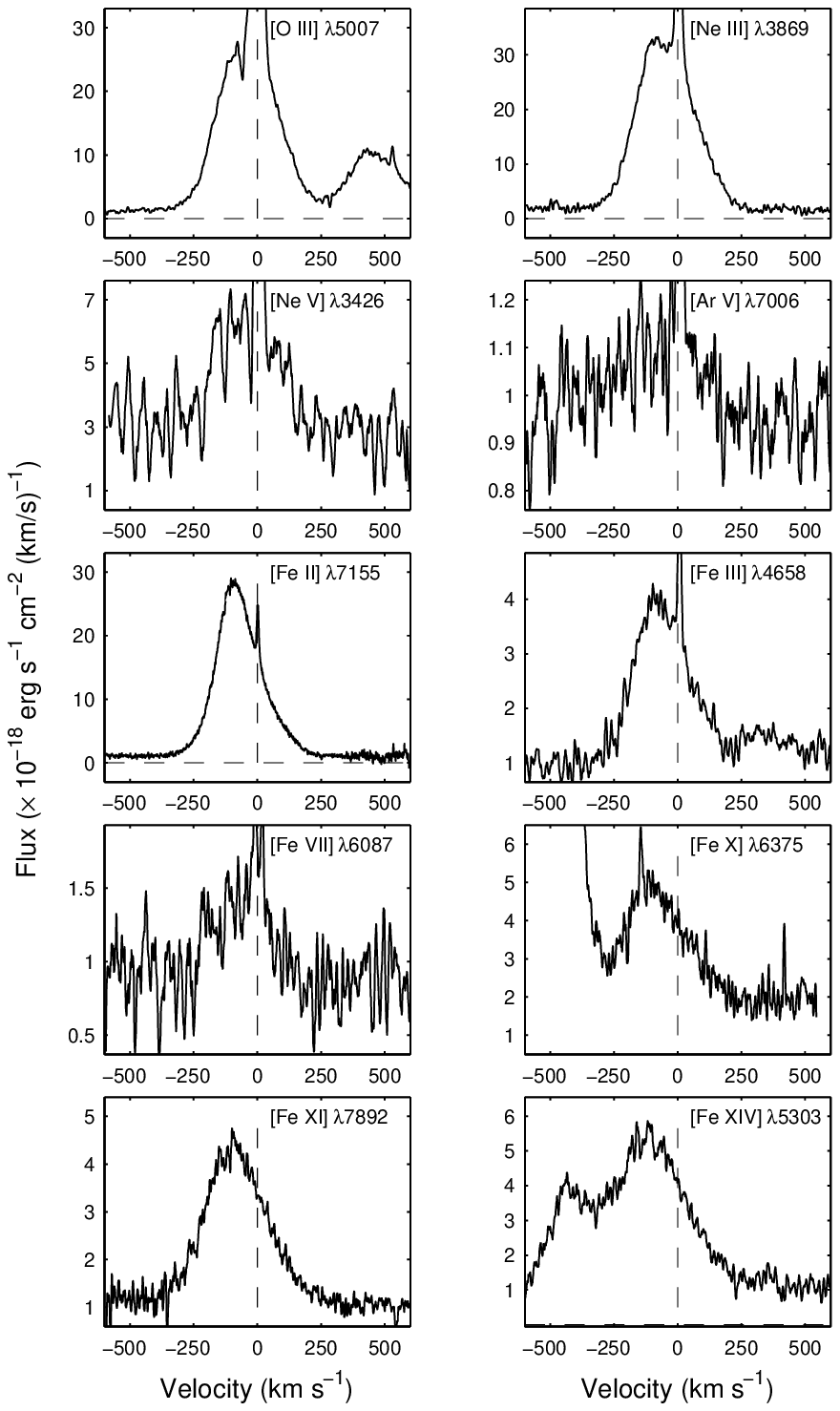}}
\end{center}
\caption{ Same as Fig. \ref{fig:high_ion_raw0012}, but for November 2005.}
\label{fig:high_ion_raw0511}
\end{figure}

In Figs. \ref{fig:high_ion_raw0012} -- \ref{fig:high_ion_raw0511} we
show a compilation of the most important of these lines for the
December 2000, October 2002, March/April 2005 and November 2005
epochs. These include [Ne~V]$\wl 3425.9$, [Ar~V]$\wl 7005.7$,
[Fe~VII]$\wl 6087.0$, [Fe~X]$\wl 6374.5$, [Fe~XI]$\wl 7891.8$, and
[Fe~XIV] $\wl 5302.9$, which are all likely to originate from the
cooling post-shock gas. In addition, we include for comparison the
[O~III]$\wl 5006.8$, [Ne~III]$\wl 3868.8$, [Fe~II]$\wl 7155.2$,
[Fe~III]$\wl 4658.0$ lines, which are, however, likely to come from
considerably cooler, photoionized gas (Sect. \ref{sect_origin}).
Because these are very useful as diagnostics of the shocked ring gas
we concentrate in this paper on these lines, and refer to \cite{gro06}
for a more detailed analysis of the full spectrum.


\begin{figure}[h]
\begin{center}
\resizebox{\hsize}{!}{\includegraphics{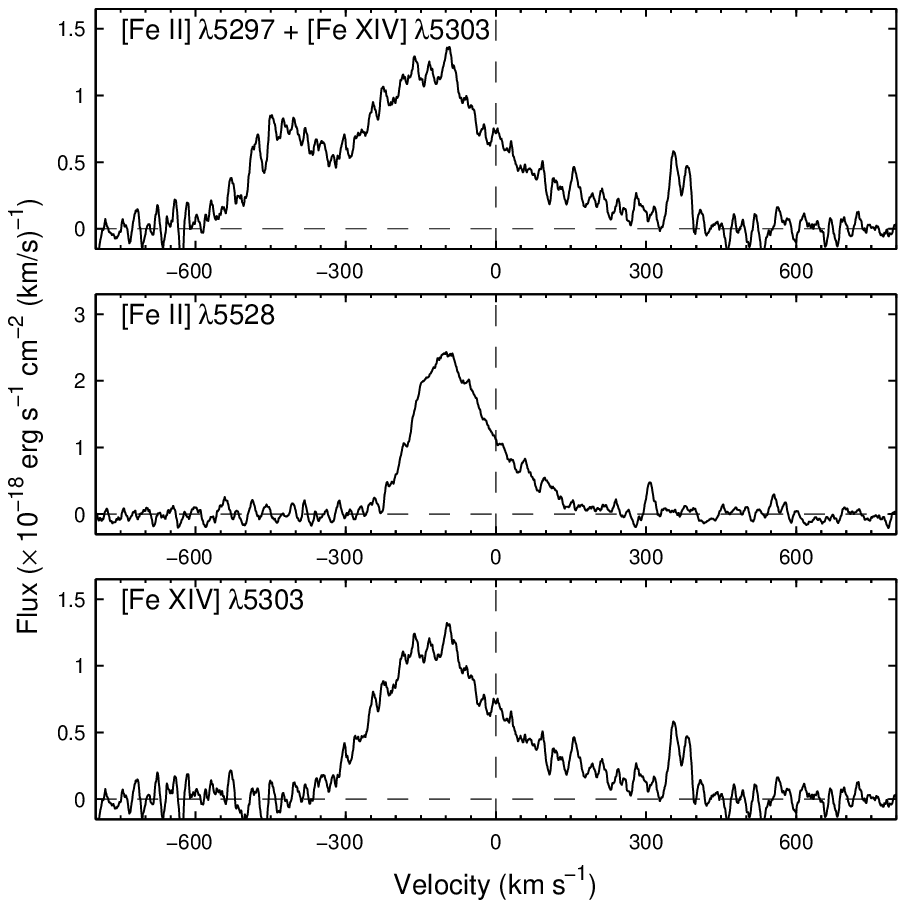}}
\end{center}
\caption{ Top panel: Line profile of [Fe~XIV]$\wl 5302.9$, before
  debelnding.
Middle panel: Template profile for [Fe~II]$\wl 5296.8$, based on the
[Fe~II]$\wl 5527.6$ line. 
Bottom panel: Deblended profile. The narrow line at $\sim 350 \kms$
in the top and bottom panels is [Ca~V]$\wl 5309.1$.}
\label{fig:fexiv_subtr}
\end{figure}

As an important example we show in
Fig. \ref{fig:fexiv_subtr} the region of the [Fe~XIV]$\wl 5302.9$
line. It is here seen that the blue wing of the [Fe~XIV]$\wl 5302.9$
line is blended with the intermediate velocity [Fe~II]$\wl 5296.8$
line. To remove this we use the [Fe~II]$\wl 5527.6$ line as a
template. The reason why we have chosen this line instead of the
stronger [Fe~II]$\wl 7155.2$ line is the similar excitation potential
to the [Fe~II]$\wl 5296.8$ line.  The
result of this subtraction can be seen in the lower panel of
Fig. \ref{fig:fexiv_subtr}. In addition, we also see a narrow
component of [Ca~V]$\wl 5309.1$ at $350 \kms$.  

The [Fe~X]$\wl 6374.5$ line is close to the red wing of the very 
strong [O~I]$\wl 6363.8$ line. In addition, there is a narrow 
Si~II$\wl 6371.4$ line close to
its peak. The [O~I]$\wl 6300.3$ line was used as a template to deblend
[O~I]$\wl 6363.8$ from the [Fe~X]$\wl 6374.5$ line. The extension of
the blue wing is therefore for this line uncertain.

\begin{figure}[h]
\begin{center}
\resizebox{\hsize}{!}{\includegraphics{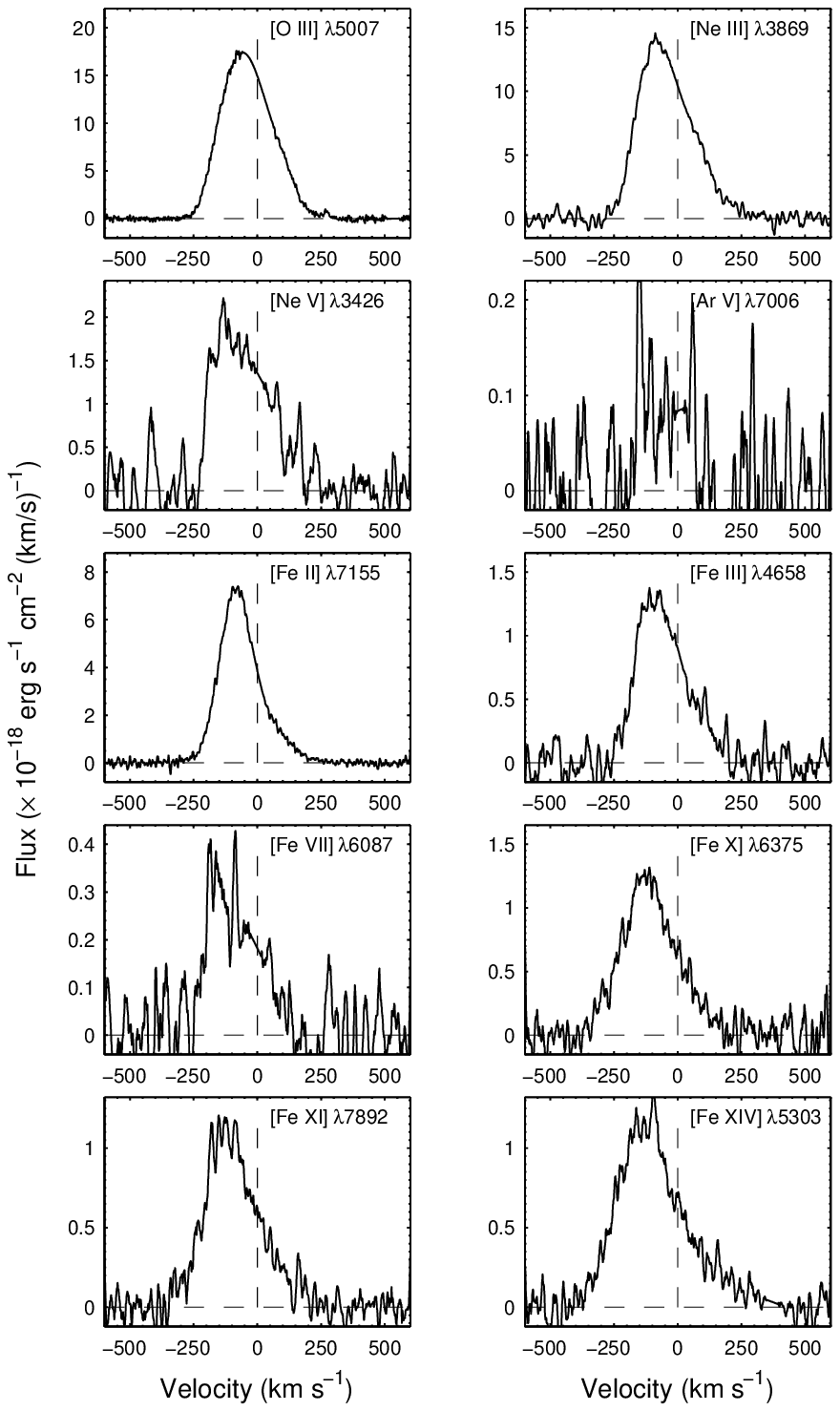}}
\end{center}
\caption{ Deblended line profiles from October 2002 of the high
  ionization lines in Fig. \ref{fig:high_ion_raw0210}}
\label{fig:high_ion}
\end{figure}

In Fig. \ref{fig:high_ion} we show the most important high ionization
lines from the October 2002 observations after deblending. The
other epochs have been deblended in the same manner. 

We have also searched the spectra for [S~XII]$\wl 7611.0$, 
[Ar~XIV]$\wl 4412.2$ and [Ca~XV]$\wll 5446.4, 5694.4$. The [S~XII] line
unfortunately falls on top of some very strong atmospheric bands. 
While in the data there is some
evidence that a line may be present at the wavelength of
the [S XII] line we cautiously attribute this to the effects
of the atmosphere.  We have also searched HST
spectra in this spectral region. Because of the low dispersion, $4.92$~\AA\
pixel$^{-1}$ or $194 \kms$ pixel$^{-1}$, and fairly low S/N, only a weak upper 
limit of $\sim 10^{-15} \ergs$ for the flux around our epoch 3 can be set.

In the region of the [Ar~XIV]$\wl 4412.2$ line there is a line
redshifted from this wavelength by $\sim 200-300 \kms$ in our VLT
spectra. This is, however, likely to be a blend of [Fe~II]$\wll 4
413.8, 4416.3$, which matches well with the velocity of the observed
feature.  Subtracting these lines, using [Fe~II]$\wl 7155.2$ as a
template in the best S/N spectrum of this region at epoch 5, we
estimate an upper limit of $\sim 15 \%$ of the [Fe~XIV] flux at this
epoch, i.e. $\la 2.2 \times 10^{-16} \ergs cm^{-2}$.

We have also looked for the [Ca~XV]$\wl 5694.4$ line, but do not detect this
to a flux limit of $\sim 15 \%$ of [Fe~XIV]$\wl 5302.9$ at epoch 4.

In Table \ref{tab:obsflux} we give the fluxes of the intermediate
velocity component for the highly ionized lines in the spectrum at the
four well observed epochs. These fluxes are corrected for blending, as
described above, but not for reddening. The estimated extinction
correction factors are listed in table \ref{tab:obsflux}. In addition,
we detect the [Fe~VI] $\wl 5335.2$ line at epoch 4 with a flux of
$(1.3 \pm 0.2)\EE{-16} \ergs cm^{-2}$.

\begin{figure}[h]
\begin{center}
\resizebox{\hsize}{!}{\includegraphics{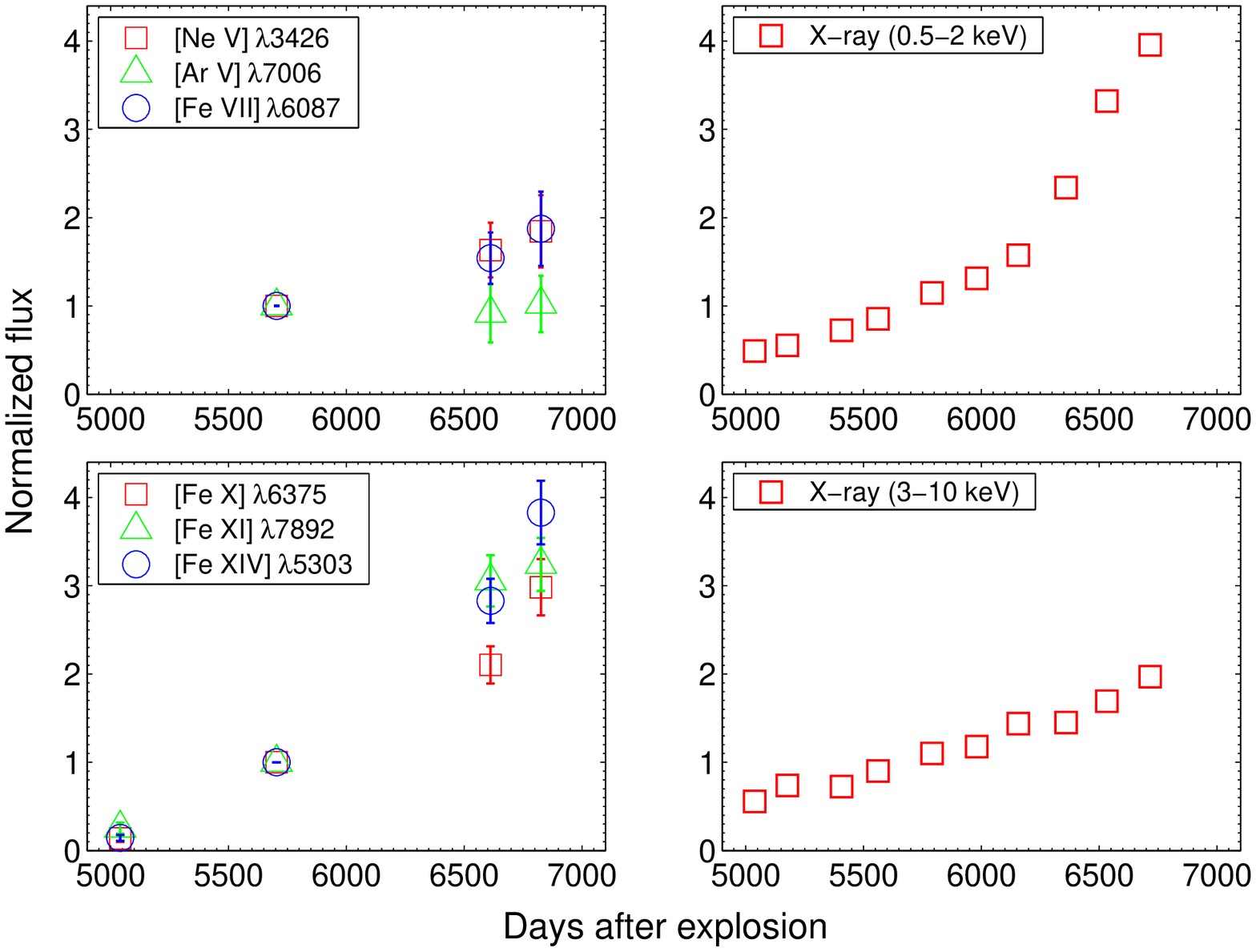}}
\end{center}
\caption{ Evolution of fluxes of the high ionization lines compared to that
of the X-ray flux from \citet{Park05}.}
\label{fig:flux_vs_time_pg}
\end{figure}

As we see from Table \ref{tab:obsflux} and
Figs. \ref{fig:high_ion_raw0012} -- \ref{fig:high_ion_raw0511}, there
is an increase in the fluxes of all lines between epochs 2 and 5.
Over this period the [Fe~X], [Fe~XI], and [Fe~XIV] lines have
increased by a factor $\sim 24-30$.  At epoch 2 some of the weaker
lines are, however, close to the noise level, and only the fluxes at
epoch 3 are reliable.  In Fig. \ref{fig:flux_vs_time_pg} we show the
evolution of the fluxes at the observed epochs.  To compare with the
X-ray and radio flux evolution, the fluxes in the figure have been
normalized to the October 2002 level, i.e., epoch 3. From this Figure
it is seen that the [Fe~X], [Fe~XI] and [Fe~XIV] lines evolve very
similar to the flux evolution of the soft X-rays, and considerably
faster than the hard X-rays or the radio. The [Ne~V], [Ar~V], and
[Fe~VII] lines, however, increase considerably more slowly.  We return to
this result in Sect. \ref{sect_disc}.

Although there seem to be considerable differences between the line
profiles of the different ions, it is important to realize that both
the deblending procedure and the low S/N for the faintest lines may
introduce considerable uncertainties. Only for lines of high S/N is it
meaningful to compare the line profiles directly.  The velocities of
the different lines will be discussed in detail in \citet{gro06}. Here
we only make some brief remarks about the most relevant issues.  If we
compare the red extension of [Fe~XIV]$\wl 5302.9$ with the [Fe~II]
$\wl 7155.2$ line we find evidence for an extension to $\sim 350 \kms$
for [Fe~XIV], while the [Fe~II] line extends to $\sim 250 \kms$. There
are, however, important systematics due to uncertainties in background
subtraction both in the spatial and dispersion directions. In
addition, deblending of line profiles gives rise to uncertainties
especially when the deblended line is strong compared to the line in
consideration. This makes e.g., the extent of the blue wing of
the [Fe~XIV]$\wl 5302.9$ line more uncertain than the red. With the
systematic uncertainties included we estimate that the values for
maximum velocities should be accurate to within $50\kms$.

\begin{table*}
\centering
\caption{Fluxes (uncorrected for reddening) of the high ionization lines from 
the north side of the shocked ring.}
\begin{tabular}{l c l c c c}
\hline
\hline
Ion&Rest wavel.&Epoch&Flux&Extinction\\ 
&\AA&&$10^{-16}\ergsm$&correction\\
\hline
&&&&\\
\lbrack Ne V\rbrack&3425.86&2&-&2.06\\
&&3&$\phantom05.25\pm 0.61$&\\
&&4&$\phantom08.58\pm 1.26$&\\
\smallskip
&&5&$\phantom09.69\pm 1.83$&\\

\lbrack Ar V\rbrack&7005.67&2&-&1.41\\
&&3&$\phantom00.34\pm 0.07$&\\
&&4&$\phantom00.31\pm 0.09$&\\
\smallskip
&&5&$\phantom00.35\pm 0.08$&\\
\
\lbrack Fe VII\rbrack&6087.0\phantom0&2&-&1.50\\
&&3&$\phantom00.80\pm 0.11$&\\
&&4&$\phantom01.23\pm 0.15$&\\
\smallskip
&&5&$\phantom01.50\pm 0.26$&\\
\
\lbrack Fe X\rbrack&6374.51&2&$\phantom00.51\pm 0.11$&1.47\\
&&3&$\phantom03.11\pm 0.24$&\\
&&4&$\phantom06.54\pm 0.37$&\\
\smallskip
&&5&$\phantom09.27\pm 0.69$&\\
\
\lbrack Fe XI\rbrack&7891.94&2&$\phantom00.91\pm 0.17$&1.32\\
&&3&$\phantom02.98\pm 0.25$&\\
&&4&$\phantom09.10\pm 0.40$&\\
\smallskip
&&5&$\phantom09.65\pm 0.34$&\\
\
\lbrack Fe XIV\rbrack&5302.86&2&$\phantom00.67\pm 0.12$&1.61\\
&&3&$\phantom03.89\pm 0.31$&\\
&&4&$11.00\pm 0.39$&\\
\smallskip
&&5&$14.89\pm 0.69$&\\
\hline
\label{tab:obsflux}
\end{tabular}

\end{table*}

\section{Origin of the coronal emission}
\label{sect_origin}
The intermediate component of both the [O~III]$\wl 5006.8$ and the
[Ne~V]$\wl 3425.9$, as well as several strong UV lines, like N~V
$\wll 1238.8, 1242.8$, Si~IV$\wll 1393.8, 1403.8$, C~IV$\wll 1548.2,
1550.8$, C~II$\wl 2325$ and Mg~II$\wll 2795.5, 2802.7$, were seen in
the HST spectra of \cite{pun02}. However, no coronal lines were seen
probably due to the lower spectral resolution, as well as the lower S/N.  The
[Fe~X]$\wl 6374.5$ and [Fe~XIV]$\wl 5302.9$ coronal lines have,
however, been observed in several supernova remnants, like the Cygnus
loop \citep{woodgate74}, Pup A, RCW86 \citep{lucke79} and IC433
\citep{woodgate79}.

\cite{pun02} propose that the intermediate component arises when the
blast wave hits dense protrusions from the ring.  The velocity of the
shocked gas behind the transmitted shock is given by $\sim (\rho_{\rm
H II}/\rho_{\rm spot})^{1/2} V_{\rm blast}$. Here $\rho_{\rm H II}$ is
the density of the H II region of the progenitor star. Estimates give
$\rho_{\rm H II}\sim 10^2 \ccm$ \citep{CD95}, while the density of the
hot spot should be close to that of the main ring material $\sim 10^4
\ccm$. For a blast wave velocity of $\sim 3800 \kms$
\citep{park05} the velocity of the transmitted shock should be $\sim
0.1 V_{\rm blast} \approx 300-500 \kms$. This is close to that
observed from the intermediate component.

The origin of the X-ray emission is, however, not so clear. Part of
the X-rays may arise from the blast wave propagating with a
velocity of $\sim 3\,500 \kms$ into the H II region of the progenitor,
interior to the equatorial ring.  \citet{Zhe05}, however, find a
velocity from the line profiles much smaller than that inferred from
the expansion from the size of the X-ray image \citep{Park04} and the
radio source \citep{man02,man05}, which implies an expansion rate of
$\sim 3\,000 \kms$. 

Zhekov \etal therefore propose a scenario where the blast wave
propagates with this comparatively high velocity inside the ring in a
medium which is dominated by a density characteristic of the H II
region, $\sim 10^2 \ccm$. In this scenario most of the X-ray line
emission arises where the blast wave hits the dense protrusions from
the ring, forming transmitted shocks into the dense gas. In addition to
this, \cite{borkowski97} find that there will be reflected shocks
going back from the hot spot into the shocked gas behind the blast
wave. The velocity of the shocked gas behind the reflected shock is
$\sim (\rho_{\rm H II}/\rho_{\rm blast})^{1/2} V_{\rm blast} \approx
0.5 V_{\rm blast} \approx 1\,000-1\,500 \kms$. This, and the even lower
velocity from the transmitted shocks, explains the low velocity
compared to the blast wave inferred from the X-ray line emitting
gas. The higher density behind reflected shocks, as well as the
transmitted, causes these to dominate the X-ray emission over the
blast wave. It also suggests that most of the X-ray emission should be
correlated with the optical emission from the hot spots. Because of
the lower density and higher temperature behind the reflected shocks,
these are likely to be adiabatic. The cooling gas seen in the coronal
lines, as well as in the optical/UV emission from lower ionization
stages, is therefore likely to arise in the radiative shocks in the
hot spots. We will therefore investigate this possibility more
quantitatively below.

For a gas with $x \equiv n({\rm He})/n({\rm H})= 0.25$ \citep{LF96} the
temperature behind the shock going into the hot spot is
\begin{eqnarray}
T_{\rm s}&=&2.27\EE{-9} {(1+4 x) \over (2+3 x)} V_{\rm s}^2  = \cr
&&1.5\EE6 \left({V_{\rm s}\over 300
  \kms}\right)^2 \ \rm K.
\label{eq:tshock}
\end{eqnarray}
 \citet{michael02} find that for a shock with velocity $500 \kms$
equipartition between electrons and ions takes place on a time scale
of $\sim 60 \ (n_e / 10^{4} \ccm)^{-1} $ days. Shocks with lower
velocity will equilibrate faster. In the following we therefore assume
that $T_{\rm e}=T_{\rm ion} = T_{\rm s}$.

For $100 \la V_{\rm s} \la 600 \kms$ the cooling function for the
CNO-enriched composition found for the ring in \cite{LF96} is
$\Lambda(T_{\rm e}) \approx 4.0\EE{-23} (T_{\rm e}/10^6 ~{\rm
K})^{-0.7} \ergs cm^{3}$, using the code by \citet{NFK05}. 
The cooling time of the shock may therefore be approximated by
\begin{equation}
t_{\mathrm{cool}} \approx 8.3 \left({n_{\rm spot} \over 10^4 \ccm}\right)^{-1} 
\left({V_{\rm s} \over 300 \kms}\right)^{3.4}
\rm
years,
\label{eq:tcool}
\end{equation}
where $n_{\rm spot}$ is the pre-shock density. Therefore, depending on
the density and the shock velocity, the shock may be either adiabatic
or radiative. While coronal emission lines may arise in any hot gas
with a temperature of $\sim 10^6$ K, the wide range of ionization
stages with similar line profiles seen in the intermediate velocity
component, from Fe~II to Fe~XIV, strongly indicates that at least some
of the emission is originating in radiative shocks.

The structure of the radiative shocks was discussed in
\cite{pun02}. Pun \etal, however, mainly focused on low and
intermediate ionization lines. To calculate the flux of the
optical/UV/IR coronal lines, as well as some of the other high
ionization lines, we compute the spectrum of the shock, using the
shock code of \cite{NFK05}. Although originally used for the
reverse shock in supernovae, this is, of course, applicable for any
radiative shock, independent of shock direction. This determines the
temperature and ionization structure of the shocked gas by solving the
time dependent ionization and hydrodynamic equations, assuming that
the structure is stationary and one-dimensional. The latter should be
sufficient as long as the shock is radiative. Collisional ionization
rates, as well as recombination rates, are from the most recent data,
and references are given in \cite{NFK05}. In addition, we include
charge transfer, which becomes important at these comparatively low
temperatures. The uncertainties of the ionization and recombination
data especially for iron have been discussed by \citet{masai97} and
\citet{gianetti00}. The emissivity is calculated using full
multi-level atoms. Pre-ionization of the unshocked gas by X-rays from
the shock is calculated using an updated version of the code in
\citet{CF94}. The abundances used are the ones used by
\citet{pun02}, which are the CNO enriched abundances for He, C, N and
O, taken from \citet{LF96}, and LMC abundances from \citet{rudop92}
for the other elements, with exception to Si which is from \citet{welty99}.

Because the focus in this paper is on the coronal lines we do not
attempt to calculate the structure of the photoionization heated gas
behind the cooling, shocked gas. This region dominates the flux for
medium and low ionization lines, like [O~III], [Ne~III] and lower
ionization stages. The structure of this region, as well as the
pre-ionized emission lines, will be discussed in a separate paper.
Here we only note that the low and intermediate ionization lines, like
[O~III], [Ne~V], [Ar~V], and [Fe~VII], all have narrow components at
zero velocity. The higher ionization lines, [Fe~X], [Fe~XI], and [Fe~XIV], 
however, lack such a component. This indicates that
pre-ionization of the unshocked gas does not reach more than the
former stages. Models similar to those in \citet{LF96} show that gas
ionized by the initial UV/soft X-ray flash at shock break out with
densities $\ga 10^4 \ccm$ is at these epochs mainly found in Fe~I-II, 
with $\la 10^{-3}$ in Fe~III. Only in gas with densities $\la 10^3 \ccm$ 
can ions like Fe~VII be found.

The collision strengths of the important coronal lines are mainly from
the Chianti data base \citep{dere97,landi06}, 
but in some cases supplemented with more recent
calculations. The critical densities for collisional de-excitation of
the [Fe~X-XIV] lines are in the range $10^7-10^9 \ccm$, and
collisional destruction is therefore not important for
these. Consequently, the relative line ratios are not sensitive to the
exact value of the density. Here we take a value of $n_{\rm spot}=10^4 \ccm$
for the pre-shock density.

\begin{figure}[t]
\begin{center}
\resizebox{\hsize}{!}{\includegraphics{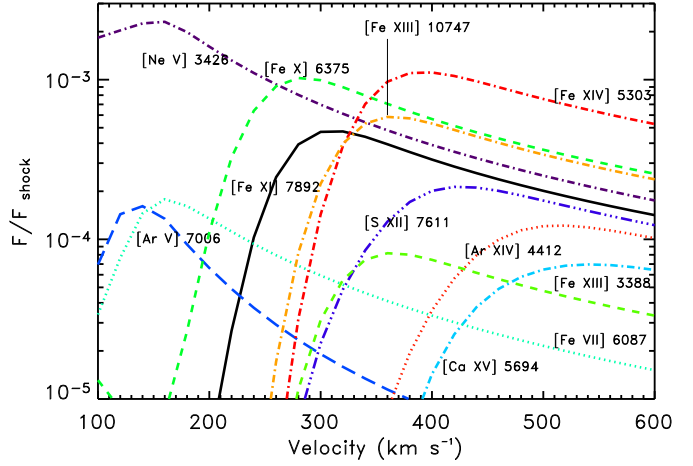}}
\end{center}
\caption{ Fluxes of the most important optical lines from radiative
shocks as function of the shock velocity . The fluxes are given
relative to the energy flux, $F_{\rm shock} = \rho V_{\rm shock}^3/2$,
through the shock.}
\label{fig:line_ratios}
\end{figure}

In Fig. \ref{fig:line_ratios} we show the strengths of the most
important optical, high ionization lines as a function of the shock
velocity relative to the kinetic flux through the shock, $\rho
V_{\rm s}^3/2$. From this we see that at shock velocities $\la 400 \kms$ 
the line ratios vary strongly with velocity, while above this velocity
they are, with the exceptions of [S~XII]$\wl 7611.0$ and 
[Ar~XIV]$\wl 4412.2$, nearly constant. This is explained by the fact that
above $\sim 400 \kms$ the most abundant ions in the immediate
post-shock gas are ions more highly ionized than Ne~V and Fe~XIV. For
a radiative shock with velocity higher than $\sim 400 \kms$ the
fraction of the cooling region above $\sim 2\EE6$ K, which corresponds
to ionization stages higher than these ions, is largely irrelevant to
the flux of the coronal lines, and the line ratios will therefore be
insensitive to the shock velocity.

\begin{figure}[h]
\begin{center}
\resizebox{\hsize}{!}{\includegraphics[angle=0]{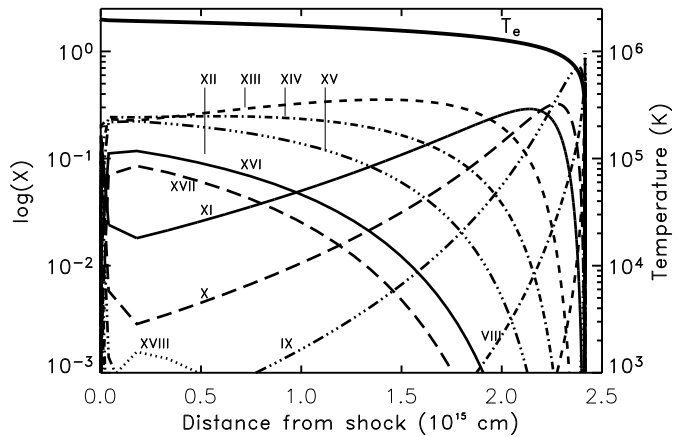}}
\end{center}
\caption{ Ionization structure for iron and temperature behind a
  radiative shock with velocity $350 \kms$ and pre-shock density $10^4
  \ccm$.}
\label{fig:shock_structure}
\end{figure}

In Fig. \ref{fig:shock_structure} we show the ionization and
temperature structure of iron of a radiative shock with velocity $350
\kms$. Immediately behind the shock in a region too small to be well
resolved on this scale, $\sim 5 \EE{13}$ cm, the ions adjust from
their pre-ionization values to their equilibrium abundances.  Behind
this region Fe~IX-XIV all have high abundances in most of the shocked
gas. As the gas cools the temperature and ionization slowly
decreases.  When the temperature has fallen below $\sim 10^6$ K
cooling increases catastrophically and drops to $(1-3) \times 10^4$
K. This region is too thin to be resolved in the plot, but is very
important for the observed emission lines (see below). The different
lines are therefore direct probes of the temperature interval where
they have their maximum abundance.

\begin{figure}    
\begin{center}
\resizebox{\hsize}{!}{\includegraphics[angle=0]{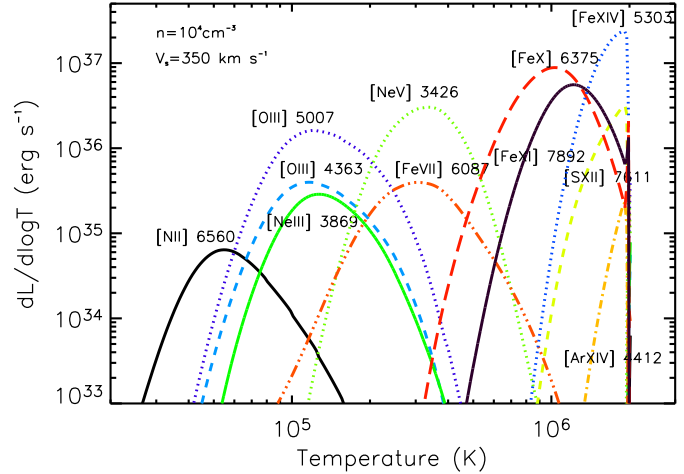}}
\end{center}
\caption{ Contribution to the luminosity per logarithmic temperature
  interval to different lines behind a shock with velocity $350 \kms$.}
\label{fig:shock_dldlogt}
\end{figure}

To more clearly illustrate the sensitivity of the different lines to the
temperature we show in Fig. \ref{fig:shock_dldlogt} the contribution
to the luminosity per logarithmic temperature interval of each line as
function of temperature. This clearly shows how the different lines
trace different temperature ranges behind the shock, with the higher
ionization stages naturally more sensitive to the higher
temperatures. Consequently, the [Fe~XI-XIV] lines arise at
$(1-2)\times 10^6$ K, while the [Ne~V], [Ar~V] and [Fe~VII] lines
arise at $\sim 3 \times 10^5$ K in a geometrically narrow, but
important, region, where the temperature falls from $10^6$ K to $10^4$
K. Because the whole post-shock region is in near pressure
equilibrium, the density is $n_e(T_e) \approx 4 n_{\rm spot} T_{\rm
s}/T_e$.  The [Ne~V] and [Fe~VII] lines e.g., arise at $T_e \sim
3\EE5$ K, corresponding to $\sim 24 n_{\rm spot} \approx 2.4 \EE5
\ccm$, for a pre-shock density of $n_{\rm spot} = 10^4
\ccm$. The critical densities for these lines are $\ga 10^7 \ccm$, and
collisional de-excitation is therefore clearly not important
for these. 

For completeness Fig. \ref{fig:shock_dldlogt} also shows the [N~II],
[O~III] and [Ne~III] lines.  In the collisionally ionized part of
the shock they each trace different temperature intervals from $\sim
10^4$ K to $\sim 4\times 10^5$ K. They all have, however, more
important contributions from the photoionized, cooling gas, not
included in these models.

To compare the models with our observations we relate all line fluxes
to that of [Fe~XIV]$\wl 5302.9$ at the different epochs. These ratios
are given for the different epochs in Table \ref{tab:relflux}. In the
same Table we also give the shock velocity corresponding to the
different line ratios, taken from Fig.  \ref{fig:line_ratios}. As can
be seen, the range of shock velocities is for all line ratios
surprisingly small, $310 - 390 \kms$. The corresponding temperature
behind the shock is $(1.6-2.5)\EE6$ K. This comparison assumes
that the emitting area in each line is the same. If there are shocks
with a range of velocities this assumption may be questionable. Because
of the rapid variation of the line ratios with velocity, this is,
however, not likely to affect the derived velocities by a large
factor.

There is some tendency for lines from ions with lower ionization
potential, e.g., Ne~V, Ar~V and Fe~VII, to give a somewhat lower
velocity than the lines from high ionization ions like Fe~X and Fe~XI.
The fact that we do not see [Ar~XIV]$\wl 4412.2$, at a flux level of
$\la 15 \%$ of the [Fe~XIV] line is consistent with a shock velocity
$\la 500 \kms$. The [Ca~XV]$\wl 5694.4$ line gives a similar upper
limit of $\sim 500 \kms$ (Fig. \ref{fig:line_ratios}).

Finally, we note that although only barely covered by these
observations, one also expects the [Fe~XIII] $\wl 3388.5$ and $\wll
10746.8 - 10797.9$ lines to be detectable in the UV and near-IR,
respectively. In particular, the IR lines are expected to have a
strength comparable to that of [Fe~X] $\wl 6374.5$ (see
Fig. \ref{fig:line_ratios}). The $\wl 3388.5$ line is within our
observed range, but the spectrum is here noisy due to the decreasing
sensitivity of the instrument and the influence of the atmosphere. We
estimate a maximum strength comparable to that of [Fe~XIV]$\wl
5302.9$. Because the expected flux is $\sim 10\%$ of this line, this
limit is not very useful. The IR lines are outside the range.

\begin{table*}
\centering
\caption{Reddening corrected fluxes relative to [Fe~XIV] $\wl 5303$
  and corresponding shock velocities ($\kms$).}
\begin{tabular}{l r r r r}
\hline
\hline
 Line&Epoch 2&Epoch 3&Epoch 4&Epoch 5 \\ 
\hline
&&&&\\
\lbrack Fe VII\rbrack $\wl 6087.0$&--&0.19&0.10&0.094\\
&--&316&330&332\\
&&&&\\
\lbrack Fe X\rbrack $\wl 6374.5$&0.65&0.68&0.54&0.57\\
&369&366&391&381\\
&&&&\\
\lbrack Fe XI\rbrack $\wl 7892.0$&1.05&0.63&0.68&0.53\\
&326&340&338&348\\
&&&&\\
\lbrack Ne V\rbrack $\wl 3425.9$&--&1.74&1.00&0.83\\
&--&310&318&321\\
&&&&\\
\lbrack Ar V\rbrack $\wl 7005.7$&--&0.076&0.025&0.021\\
&--&313&336&339\\
&&&&\\
\hline
\label{tab:relflux}
\end{tabular}

\end{table*}

\section{Discussion}
\label{sect_disc}

Although we get good agreement for a rather narrow range of shock
velocities, it is clear that the dynamics is more complex.  This can
be seen in the hydrodynamic simulations of the collision performed by
several groups \citep{Luo94,Masai94,borkowski97}.  In particular,
\citet{borkowski97} have modeled the structure and X-ray emission from
the impact with 2-D hydro simulations. The interplay of the shock
propagating into the dense ring gas and reflected shocks from the
ejecta--CSM gives a time dependence, as well as spatial dependence,
which is complex and only can be described by detailed
multi-dimensional hydrodynamical simulations. Although the Borkowski
et. al. simulations used a simplified model for the ring structure, they
give a good idea of the processes and complications involved in
modelling the emission.

A further complication is that the observed spectra are a convolution
of several hot spots seen with different projected velocities. In
addition, the number of spots is increasing. In these respects our
observations have similar limitations as the Chandra X-rays
observations, although our spectral resolution is much
better. Spatially resolved spectroscopy would be of obvious
importance. 

In Sect. 3 we found that the [Fe~XIV] $\wl 5302.9$ line extends to
$\sim 350 \kms$, while the lower ionization lines only extend to $\sim
250 \kms$. This may indicate that the low ionization lines have a
dominant contribution from tangential shocks with lower velocity,
while the high ionization lines also come from shocks perpendicular
to the blast wave, having a high velocity. 

\cite{michael02} find that the relative fluxes of the lines observed
with the grating spectrometer on Chandra in October 1999 are best
fitted with a shock temperature of $k T_e \sim 2.9 \pm 0.4$ keV. The
lower dispersion CCD spectra obtained in December 2000 indicated a
marginally lower temperature, $k T_e \sim 2.7 \pm 0.2$ keV. In
addition, they find that they get a slightly better fit to the
spectrum with an additional softer component with $k T_e \sim 0.4 \pm
0.1$ keV. The flux contribution from this was $\sim 7 \%$. It is worth
noting that even the high temperature component has an electron
temperature that is considerably lower than the shock temperature,
corresponding to equipartition, $\sim 17$ keV. This is in common with
several other SNRs.

Due to the increase in the X-ray flux \cite{Zhe05,zhe06} were able to
repeat this study with considerably higher S/N.  Zhekov et. al. find
that they get a good fit to the X-ray line ratios, as well as the
widths, for a model with a range of temperature, $0.15-4$ keV,
corresponding to $V_s = 340 - 1\,700 \kms$. The lower part of this range
is similar to that observed in the optical lines, and it is very
likely that they arise from the same region. As \cite{Zhe05} remark,
the upper range of velocities is, however, likely to come from
adiabatic shocks, probably reflected from the impact on the
protrusions. If they are adiabatic and with velocity $\ga 400 \kms$,
corresponding to a temperature of $\ga 3\EE6$ K, they will not produce
optical emission.

We propose that the regions that produce the soft X-ray flux also
account for most of the emission in the [Fe~X], [Fe~XI] and [Fe~XIV]
lines. This is supported by the similar evolution of the flux between
these two components (Fig. \ref{fig:flux_vs_time_pg}), and, as
mentioned above, the comparable shock velocities required to produce
them, $\sim 350 \kms$ (see also Table 3). However, in order to explain
the slower evolution of the intermediate ionization lines such as [Ne
V], [Ar~V] and [Fe~VII], we do require that some fraction of the
[Fe~X-XIV] fluxes arise from adiabatic shocks which would not be
contributing to the intermediate and low ionization lines. The
sensitivity of the cooling time scales to the shock velocities and the
density of the hot spots (Eq. \ref{eq:tcool}) makes this a plausible
scenario. The somewhat narrower line widths that the intermediate
ionization lines exhibit also support this argumentation.

\section{Conclusions and summary}
\label{sect_concl}

The most important result of this paper is the discovery of a number
of high ionization lines coming from gas of temperatures up to $\sim
2\EE6$ K. The rapidly increasing flux of these is well correlated to
the flux in the soft X-rays, and offer a complimentary view of the
interaction of the ejecta and the ring material. The large range in
ionization stage with similar line profiles shows that most of the
emission is coming from radiative shocks with velocity $310-390 \kms$,
although a fraction of the coronal line emission may originate in
adiabatic shocks. The shock velocity we find from our spectral
modeling is consistent with the width of the lines.  We hope in the
future to do a more detailed modeling of the spectra, including both
the lower ionization lines from the photoionized, cooling gas, as well
as the X-ray lines. Also a more detailed modeling of the hydrodynamics
of the explosion along the lines of \cite{borkowski97} with these extra
constraints from the line strengths, as well as the line profiles
would be highly interesting. Continued monitoring of the ring collision
is of obvious high importance. The increasing flux, as well as new
instruments using adaptive optics, will enable us to study this unique
collision in even more detail.


\begin{acknowledgements}
We are grateful to the referee for several useful comments. This work
was supported in part by the Swedish Research Council and the Swedish
National Space Board. Part of this research was performed while CF and
RAC were visiting the Kavli Institute of Theoretical Physics,
supported in part by the National Science Foundation under Grant
No. PHY99-07949. PL is a Research Fellow at the Royal Swedish Academy
supported by a grant from the Wallenberg Foundation.
\end{acknowledgements}


\end{document}